
%
\catcode`@=11 
%
%
%

\font\fourteenrm=cmr10 scaled\magstep2
\font\twelverm=cmr10 scaled\magstep1
\font\ninerm=cmr9            \font\sixrm=cmr6

\font\fourteenbf=cmbx10 scaled\magstep2
\font\twelvebf=cmbx10 scaled\magstep1
\font\ninebf=cmbx9            \font\sixbf=cmbx6
\font\seventeeni=cmmi10 scaled\magstep3     \skewchar\seventeeni='177
\font\fourteeni=cmmi10 scaled\magstep2      \skewchar\fourteeni='177
\font\twelvei=cmmi10 scaled\magstep1        \skewchar\twelvei='177
\font\ninei=cmmi9                           \skewchar\ninei='177
\font\sixi=cmmi6                            \skewchar\sixi='177
\font\seventeensy=cmsy10 scaled\magstep3    \skewchar\seventeensy='60
\font\fourteensy=cmsy10 scaled\magstep2     \skewchar\fourteensy='60
\font\twelvesy=cmsy10 scaled\magstep1       \skewchar\twelvesy='60
\font\ninesy=cmsy9                          \skewchar\ninesy='60
\font\sixsy=cmsy6                           \skewchar\sixsy='60

\font\fourteenex=cmex10 scaled\magstep2
\font\twelveex=cmex10 scaled\magstep1

\font\fourteensl=cmsl10 scaled\magstep2
\font\twelvesl=cmsl10 scaled\magstep1
\font\ninesl=cmsl9

\font\fourteenit=cmti10 scaled\magstep2
\font\twelveit=cmti10 scaled\magstep1
\font\twelvett=cmtt10 scaled\magstep1
\font\twelvecp=cmcsc10 scaled\magstep1
\font\tencp=cmcsc10
\newfam\cpfam
%
%
\newcount\f@ntkey            \f@ntkey=0
\def\samef@nt{\relax \ifcase\f@ntkey \rm \or\oldstyle \or\or
         \or\it \or\sl \or\bf \or\tt \or\caps \fi }
\def\fourteenpoint{\relax
    \textfont0=\fourteenrm          \scriptfont0=\tenrm
    \scriptscriptfont0=\sevenrm
     \def\rm{\fam0 \fourteenrm \f@ntkey=0 }\relax
    \textfont1=\fourteeni           \scriptfont1=\teni
    \scriptscriptfont1=\seveni
     \def\oldstyle{\fam1 \fourteeni\f@ntkey=1 }\relax
    \textfont2=\fourteensy          \scriptfont2=\tensy
    \scriptscriptfont2=\sevensy
    \textfont3=\fourteenex     \scriptfont3=\fourteenex
    \scriptscriptfont3=\fourteenex
    \def\it{\fam\itfam \fourteenit\f@ntkey=4 }\textfont\itfam=\fourteenit
    \def\sl{\fam\slfam \fourteensl\f@ntkey=5 }\textfont\slfam=\fourteensl
    \scriptfont\slfam=\tensl
    \def\bf{\fam\bffam \fourteenbf\f@ntkey=6 }\textfont\bffam=\fourteenbf
    \scriptfont\bffam=\tenbf     \scriptscriptfont\bffam=\sevenbf
    \def\tt{\fam\ttfam \twelvett \f@ntkey=7 }\textfont\ttfam=\twelvett
    \h@big=11.9\p@{} \h@Big=16.1\p@{} \h@bigg=20.3\p@{} \h@Bigg=24.5\p@{}
    \def\caps{\fam\cpfam \twelvecp \f@ntkey=8 }\textfont\cpfam=\twelvecp
    \setbox\strutbox=\hbox{\vrule height 12pt depth 5pt width\z@}
    \samef@nt}
\def\twelvepoint{\relax
    \textfont0=\twelverm          \scriptfont0=\ninerm
    \scriptscriptfont0=\sixrm
     \def\rm{\fam0 \twelverm \f@ntkey=0 }\relax
    \textfont1=\twelvei           \scriptfont1=\ninei
    \scriptscriptfont1=\sixi
     \def\oldstyle{\fam1 \twelvei\f@ntkey=1 }\relax
    \textfont2=\twelvesy          \scriptfont2=\ninesy
    \scriptscriptfont2=\sixsy
    \textfont3=\twelveex          \scriptfont3=\twelveex
    \scriptscriptfont3=\twelveex
    \def\it{\fam\itfam \twelveit \f@ntkey=4 }\textfont\itfam=\twelveit
    \def\sl{\fam\slfam \twelvesl \f@ntkey=5 }\textfont\slfam=\twelvesl
    \scriptfont\slfam=\ninesl
    \def\bf{\fam\bffam \twelvebf \f@ntkey=6 }\textfont\bffam=\twelvebf
    \scriptfont\bffam=\ninebf     \scriptscriptfont\bffam=\sixbf
    \def\tt{\fam\ttfam \twelvett \f@ntkey=7 }\textfont\ttfam=\twelvett
    \h@big=10.2\p@{}
    \h@Big=13.8\p@{}
    \h@bigg=17.4\p@{}
    \h@Bigg=21.0\p@{}
    \def\caps{\fam\cpfam \twelvecp \f@ntkey=8 }\textfont\cpfam=\twelvecp
    \setbox\strutbox=\hbox{\vrule height 10pt depth 4pt width\z@}
    \samef@nt}
\def\tenpoint{\relax
    \textfont0=\tenrm          \scriptfont0=\sevenrm
    \scriptscriptfont0=\fiverm
    \def\rm{\fam0 \tenrm \f@ntkey=0 }\relax
    \textfont1=\teni           \scriptfont1=\seveni
    \scriptscriptfont1=\fivei
    \def\oldstyle{\fam1 \teni \f@ntkey=1 }\relax
    \textfont2=\tensy          \scriptfont2=\sevensy
    \scriptscriptfont2=\fivesy
    \textfont3=\tenex          \scriptfont3=\tenex
    \scriptscriptfont3=\tenex
    \def\it{\fam\itfam \tenit \f@ntkey=4 }\textfont\itfam=\tenit
    \def\sl{\fam\slfam \tensl \f@ntkey=5 }\textfont\slfam=\tensl
    \def\bf{\fam\bffam \tenbf \f@ntkey=6 }\textfont\bffam=\tenbf
    \scriptfont\bffam=\sevenbf     \scriptscriptfont\bffam=\fivebf
    \def\tt{\fam\ttfam \tentt \f@ntkey=7 }\textfont\ttfam=\tentt
    \def\caps{\fam\cpfam \tencp \f@ntkey=8 }\textfont\cpfam=\tencp
    \setbox\strutbox=\hbox{\vrule height 8.5pt depth 3.5pt width\z@}
    \samef@nt}
%
%
%
%
\newdimen\h@big  \h@big=8.5\p@
\newdimen\h@Big  \h@Big=11.5\p@
\newdimen\h@bigg  \h@bigg=14.5\p@
\newdimen\h@Bigg  \h@Bigg=17.5\p@
\def\big#1{{\hbox{$\left#1\vbox to\h@big{}\right.\n@space$}}}
\def\Big#1{{\hbox{$\left#1\vbox to\h@Big{}\right.\n@space$}}}
\def\bigg#1{{\hbox{$\left#1\vbox to\h@bigg{}\right.\n@space$}}}
\def\Bigg#1{{\hbox{$\left#1\vbox to\h@Bigg{}\right.\n@space$}}}
%
%
%
\normalbaselineskip = 20pt plus 0.2pt minus 0.1pt
\normallineskip = 1.5pt plus 0.1pt minus 0.1pt
\normallineskiplimit = 1.5pt
\newskip\normaldisplayskip
\normaldisplayskip = 20pt plus 5pt minus 10pt
\newskip\normaldispshortskip
\normaldispshortskip = 6pt plus 5pt
\newskip\normalparskip
\normalparskip = 6pt plus 2pt minus 1pt
\newskip\skipregister
\skipregister = 5pt plus 2pt minus 1.5pt
\newif\ifsingl@    \newif\ifdoubl@
\newif\iftwelv@    \twelv@true
\def\singlespace{\singl@true\doubl@false\spaces@t}
\def\doublespace{\singl@false\doubl@true\spaces@t}
\def\normalspace{\singl@false\doubl@false\spaces@t}
\def\Tenpoint{\tenpoint\twelv@false\spaces@t}
\def\Twelvepoint{\twelvepoint\twelv@true\spaces@t}
\def\spaces@t{\relax%
 \iftwelv@ \ifsingl@\subspaces@t3:4;\else\subspaces@t29:31;\fi%
 \else \ifsingl@\subspaces@t3:5;\else\subspaces@t4:5;\fi \fi%
 \ifdoubl@ \multiply\baselineskip by 5%
 \divide\baselineskip by 4 \fi \unskip}
\def\subspaces@t#1:#2;{
      \baselineskip = \normalbaselineskip
      \multiply\baselineskip by #1 \divide\baselineskip by #2
      \lineskip = \normallineskip
      \multiply\lineskip by #1 \divide\lineskip by #2
      \lineskiplimit = \normallineskiplimit
      \multiply\lineskiplimit by #1 \divide\lineskiplimit by #2
      \parskip = \normalparskip
      \multiply\parskip by #1 \divide\parskip by #2
      \abovedisplayskip = \normaldisplayskip
      \multiply\abovedisplayskip by #1 \divide\abovedisplayskip by #2
      \belowdisplayskip = \abovedisplayskip
      \abovedisplayshortskip = \normaldispshortskip
      \multiply\abovedisplayshortskip by #1
        \divide\abovedisplayshortskip by #2
      \belowdisplayshortskip = \abovedisplayshortskip
      \advance\belowdisplayshortskip by \belowdisplayskip
      \divide\belowdisplayshortskip by 2
      \smallskipamount = \skipregister
      \multiply\smallskipamount by #1 \divide\smallskipamount by #2
      \medskipamount = \smallskipamount \multiply\medskipamount by 2
      \bigskipamount = \smallskipamount \multiply\bigskipamount by 4 }
\def\normalbaselines{ \baselineskip=\normalbaselineskip
   \lineskip=\normallineskip \lineskiplimit=\normallineskip
   \iftwelv@\else \multiply\baselineskip by 4 \divide\baselineskip by 5
     \multiply\lineskiplimit by 4 \divide\lineskiplimit by 5
     \multiply\lineskip by 4 \divide\lineskip by 5 \fi }
\Twelvepoint  
\interlinepenalty=50
\interfootnotelinepenalty=5000
\predisplaypenalty=9000
\postdisplaypenalty=500
\hfuzz=1pt
\vfuzz=0.2pt
%
%
%
\def\pagecontents{
   \ifvoid\topins\else\unvbox\topins\vskip\skip\topins\fi
   \dimen@ = \dp255 \unvbox255
   \ifvoid\footins\else\vskip\skip\footins\footrule\unvbox\footins\fi
   \ifr@ggedbottom \kern-\dimen@ \vfil \fi }
\def\makeheadline{\vbox to 0pt{ \skip@=\topskip
      \advance\skip@ by -12pt \advance\skip@ by -2\normalbaselineskip
      \vskip\skip@ \line{\vbox to 12pt{}\the\headline} \vss
      }\nointerlineskip}
\def\makefootline{\baselineskip = 1.5\normalbaselineskip
                 \line{\the\footline}}
\newif\iffrontpage
\newif\ifletterstyle
\newif\ifp@genum
\def\nopagenumbers{\p@genumfalse}
\def\pagenumbers{\p@genumtrue}
\pagenumbers
\newtoks\paperheadline
\newtoks\letterheadline
\newtoks\letterfrontheadline
\newtoks\lettermainheadline
\newtoks\paperfootline
\newtoks\letterfootline
\newtoks\date
\footline={\ifletterstyle\the\letterfootline\else\the\paperfootline\fi}
\paperfootline={\hss\iffrontpage\else\ifp@genum\tenrm\folio\hss\fi\fi}
\letterfootline={\hfil}
\headline={\ifletterstyle\the\letterheadline\else\the\paperheadline\fi}
\paperheadline={\hfil}
\letterheadline{\iffrontpage\the\letterfrontheadline
     \else\the\lettermainheadline\fi}
\lettermainheadline={\rm\ifp@genum page \ \folio\fi\hfil\the\date}
\def\monthname{\relax\ifcase\month 0/\or January\or February\or
   March\or April\or May\or June\or July\or August\or September\or
   October\or November\or December\else\number\month/\fi}
\date={\monthname\ \number\day, \number\year}
\countdef\pagenumber=1  \pagenumber=1
\def\advancepageno{\global\advance\pageno by 1
   \ifnum\pagenumber<0 \global\advance\pagenumber by -1
    \else\global\advance\pagenumber by 1 \fi \global\frontpagefalse }
\def\folio{\ifnum\pagenumber<0 \romannumeral-\pagenumber
           \else \number\pagenumber \fi }
\def\footrule{\dimen@=\prevdepth\nointerlineskip
   \vbox to 0pt{\vskip -0.25\baselineskip \hrule width 0.35\hsize \vss}
   \prevdepth=\dimen@ }
\newtoks\foottokens
\foottokens={\Tenpoint\singlespace}
\newdimen\footindent
\footindent=24pt
\def\vfootnote#1{\insert\footins\bgroup  \the\foottokens
   \interlinepenalty=\interfootnotelinepenalty \floatingpenalty=20000
   \splittopskip=\ht\strutbox \boxmaxdepth=\dp\strutbox
   \leftskip=\footindent \rightskip=\z@skip
   \parindent=0.5\footindent \parfillskip=0pt plus 1fil
   \spaceskip=\z@skip \xspaceskip=\z@skip
   \Textindent{$ #1 $}\footstrut\futurelet\next\fo@t}
\def\Textindent#1{\noindent\llap{#1\enspace}\ignorespaces}
\def\footnote#1{\attach{#1}\vfootnote{#1}}

\def\foot{\attach\footsymbolgen\vfootnote{\footsymbol}}
\let\footsymbol=\star
\newcount\lastf@@t           \lastf@@t=-1
\newcount\footsymbolcount    \footsymbolcount=0
\newif\ifPhysRev
\def\footsymbolgen{\relax \ifPhysRev \iffrontpage \NPsymbolgen\else
      \PRsymbolgen\fi \else \NPsymbolgen\fi
   \global\lastf@@t=\pageno \footsymbol }
\def\NPsymbolgen{\ifnum\footsymbolcount<0 \global\footsymbolcount=0\fi
   {\iffrontpage \else \advance\lastf@@t by 1 \fi
    \ifnum\lastf@@t<\pageno \global\footsymbolcount=0
     \else \global\advance\footsymbolcount by 1 \fi }
   \ifcase\footsymbolcount \fd@f\star\or \fd@f\dagger\or \fd@f\ast\or
    \fd@f\ddagger\or \fd@f\natural\or \fd@f\diamond\or \fd@f\bullet\or
    \fd@f\nabla\else \fd@f\dagger\global\footsymbolcount=0 \fi }
\def\fd@f#1{\xdef\footsymbol{#1}}
\def\PRsymbolgen{\ifnum\footsymbolcount>0 \global\footsymbolcount=0\fi
      \global\advance\footsymbolcount by -1
      \xdef\footsymbol{\sharp\number-\footsymbolcount} }
\def\space@ver#1{\let\@sf=\empty \ifmmode #1\else \ifhmode
   \edef\@sf{\spacefactor=\the\spacefactor}\unskip${}#1$\relax\fi\fi}
\def\attach#1{\space@ver{\strut^{\mkern 2mu #1} }\@sf\ }
\def\atttach#1{\space@ver{\strut{\mkern 2mu #1} }\@sf\ }
%
%
%
\newcount\chapternumber      \chapternumber=0
\newcount\sectionnumber      \sectionnumber=0
\newcount\equanumber         \equanumber=0
\let\chapterlabel=0
\newtoks\chapterstyle        \chapterstyle={\Number}
\newskip\chapterskip         \chapterskip=\bigskipamount
\newskip\sectionskip         \sectionskip=\medskipamount
\newskip\headskip            \headskip=8pt plus 3pt minus 3pt
\newdimen\chapterminspace    \chapterminspace=15pc
\newdimen\sectionminspace    \sectionminspace=10pc
\newdimen\referenceminspace  \referenceminspace=25pc
\def\chapterreset{\global\advance\chapternumber by 1
   \ifnum\the\equanumber<0 \else\global\equanumber=0\fi
   \sectionnumber=0 \makel@bel}
\def\makel@bel{\xdef\chapterlabel{%
\the\chapterstyle{\the\chapternumber}.}}
\def\sectionlabel{\number\sectionnumber \quad }
\def\alphabetic#1{\count255='140 \advance\count255 by #1\char\count255}
\def\Alphabetic#1{\count255='100 \advance\count255 by #1\char\count255}
\def\Roman#1{\uppercase\expandafter{\romannumeral #1}}
\def\roman#1{\romannumeral #1}
\def\Number#1{\number #1}
\def\unnumberedchapters{\let\makel@bel=\relax \let\chapterlabel=\relax
\let\sectionlabel=\relax \equanumber=-1 }
\def\titlestyle#1{\par\begingroup \interlinepenalty=9999
     \leftskip=0.02\hsize plus 0.23\hsize minus 0.02\hsize
     \rightskip=\leftskip \parfillskip=0pt
     \hyphenpenalty=9000 \exhyphenpenalty=9000
     \tolerance=9999 \pretolerance=9000
     \spaceskip=0.333em \xspaceskip=0.5em
     \iftwelv@\fourteenpoint\else\twelvepoint\fi
   \noindent #1\par\endgroup }
\def\spacecheck#1{\dimen@=\pagegoal\advance\dimen@ by -\pagetotal
   \ifdim\dimen@<#1 \ifdim\dimen@>0pt \vfil\break \fi\fi}
\def\chapter#1{\par \penalty-300 \vskip\chapterskip
   \spacecheck\chapterminspace
   \chapterreset \titlestyle{\chapterlabel \ #1}
   \nobreak\vskip\headskip \penalty 30000
   \wlog{\string\chapter\ \chapterlabel} }

\def\section#1{\par \ifnum\the\lastpenalty=30000\else
   \penalty-200\vskip\sectionskip \spacecheck\sectionminspace\fi
   \wlog{\string\section\ \chapterlabel \the\sectionnumber}
   \global\advance\sectionnumber by 1  \noindent
   {\caps\enspace\chapterlabel \sectionlabel #1}\par
   \nobreak\vskip\headskip \penalty 30000 }
\def\subsection#1{\par
   \ifnum\the\lastpenalty=30000\else \penalty-100\smallskip \fi
   \noindent\undertext{#1}\enspace \vadjust{\penalty5000}}

\def\undertext#1{\vtop{\hbox{#1}\kern 1pt \hrule}}
\def\APPENDIX#1#2{\par\penalty-300\vskip\chapterskip
   \spacecheck\chapterminspace \chapterreset \xdef\chapterlabel{#1}
   \titlestyle{APPENDIX #2} \nobreak\vskip\headskip \penalty 30000
   \wlog{\string\Appendix\ \chapterlabel} }
\def\Appendix#1{\APPENDIX{#1}{#1}}
\def\appendix{\APPENDIX{A}{}}
%
%
%
\def\eqname#1{\relax \ifnum\the\equanumber<0
     \xdef#1{{\rm(\number-\equanumber)}}\global\advance\equanumber by -1
    \else \global\advance\equanumber by 1
      \xdef#1{{\rm(\chapterlabel \number\equanumber)}} \fi}

\def\eqn#1{\eqno\eqname{#1}#1}

\def\eqinsert#1{\noalign{\dimen@=\prevdepth \nointerlineskip
   \setbox0=\hbox to\displaywidth{\hfil #1}
   \vbox to 0pt{\vss\hbox{$\!\box0\!$}\kern-0.5\baselineskip}
   \prevdepth=\dimen@}}
\def\sequentialequations{\globaleqnumbers}
%
%
\def\GENITEM#1;#2{\par \hangafter=0 \hangindent=#1
    \Textindent{$ #2 $}\ignorespaces}
\outer\def\newitem#1=#2;{\gdef#1{\GENITEM #2;}}
\newdimen\itemsize                \itemsize=30pt
\newitem\item=1\itemsize;
\newitem\sitem=1.75\itemsize;     
\newitem\ssitem=2.5\itemsize;     
\outer\def\newlist#1=#2&#3&#4;{\toks0={#2}\toks1={#3}%
   \count255=\escapechar \escapechar=-1
   \alloc@0\list\countdef\insc@unt\listcount     \listcount=0
   \edef#1{\par
      \countdef\listcount=\the\allocationnumber
      \advance\listcount by 1
      \hangafter=0 \hangindent=#4
      \Textindent{\the\toks0{\listcount}\the\toks1}}
   \expandafter\expandafter\expandafter
    \edef\c@t#1{begin}{\par
      \countdef\listcount=\the\allocationnumber \listcount=1
      \hangafter=0 \hangindent=#4
      \Textindent{\the\toks0{\listcount}\the\toks1}}
   \expandafter\expandafter\expandafter
    \edef\c@t#1{con}{\par \hangafter=0 \hangindent=#4 \noindent}
   \escapechar=\count255}
\def\c@t#1#2{\csname\string#1#2\endcsname}
\newlist\point=\Number&.&1.0\itemsize;
\newlist\subpoint=(\alphabetic&)&1.75\itemsize;
\newlist\subsubpoint=(\roman&)&2.5\itemsize;
\newlist\cpoint=\Roman&.&1.0\itemsize;
%

%
%
%
\newcount\referencecount     \referencecount=0
\newif\ifreferenceopen       \newwrite\referencewrite
\newtoks\rw@toks
\def\NPrefmark#1{\atttach{\rm [ #1 ] }}
\let\PRrefmark=\attach
\def\CErefmark#1{\attach{\scriptstyle  #1 ) }}
\def\refmark#1{\relax\ifPhysRev\PRrefmark{#1}\else\NPrefmark{#1}\fi}
\def\crefmark#1{\relax\CErefmark{#1}}
\def\refend{\refmark{\number\referencecount}}
\newcount\lastrefsbegincount \lastrefsbegincount=0
\def\refsend{\refmark{\count255=\referencecount
   \advance\count255 by-\lastrefsbegincount
   \ifcase\count255 \number\referencecount
   \or \number\lastrefsbegincount,\number\referencecount
   \else \number\lastrefsbegincount-\number\referencecount \fi}}
\def\crefsend{\crefmark{\count255=\referencecount
   \advance\count255 by-\lastrefsbegincount
   \ifcase\count255 \number\referencecount
   \or \number\lastrefsbegincount,\number\referencecount
   \else \number\lastrefsbegincount-\number\referencecount \fi}}
\def\refch@ck{\chardef\rw@write=\referencewrite
   \ifreferenceopen \else \referenceopentrue
   \immediate\openout\referencewrite=referenc.texauxil \fi}
%
{\catcode`\^^M=\active 
  \gdef\obeyendofline{\catcode`\^^M\active \let^^M\ }}%
%
{\catcode`\^^M=\active 
  \gdef\ignoreendofline{\catcode`\^^M=5}}
{\obeyendofline\gdef\rw@start#1{\def\t@st{#1} \ifx\t@st\blankend%
\endgroup \@sf \relax \else \ifx\t@st\bl@nkend \endgroup \@sf \relax%
\else \rw@begin#1
\backtotext
\fi \fi } }
{\obeyendofline\gdef\rw@begin#1
{\def\n@xt{#1}\rw@toks={#1}\relax%
\rw@next}}
\def\blankend{}
{\obeylines\gdef\bl@nkend{
}}
\newif\iffirstrefline  \firstreflinetrue
\def\rwr@teswitch{\ifx\n@xt\blankend \let\n@xt=\rw@begin %
 \else\iffirstrefline \global\firstreflinefalse%
\immediate\write\rw@write{\noexpand\obeyendofline \the\rw@toks}%
\let\n@xt=\rw@begin%
      \else\ifx\n@xt\rw@@d \def\n@xt{\immediate\write\rw@write{%
        \noexpand\ignoreendofline}\endgroup \@sf}%
             \else \immediate\write\rw@write{\the\rw@toks}%
             \let\n@xt=\rw@begin\fi\fi \fi}
\def\rw@next{\rwr@teswitch\n@xt}
\def\rw@@d{\backtotext} \let\rw@end=\relax
\let\backtotext=\relax

\newdimen\refindent     \refindent=30pt
\def\refitem#1{\par \hangafter=0 \hangindent=\refindent \Textindent{#1}}
\def\REFNUM#1{\space@ver{}\refch@ck \firstreflinetrue%
 \global\advance\referencecount by 1 \xdef#1{\the\referencecount}}
\def\refnum#1{\space@ver{}\refch@ck \firstreflinetrue%
 \global\advance\referencecount by 1 \xdef#1{\the\referencecount}\refend}

\def\REF#1{\REFNUM#1%
 \immediate\write\referencewrite{%
 \noexpand\refitem{#1.}}%
\begingroup\obeyendofline\rw@start}
\def\ref{\refnum\?%
 \immediate\write\referencewrite{\noexpand\refitem{\?.}}%
\begingroup\obeyendofline\rw@start}
\def\Ref#1{\refnum#1%
 \immediate\write\referencewrite{\noexpand\refitem{#1.}}%
\begingroup\obeyendofline\rw@start}
\def\REFS#1{\REFNUM#1\global\lastrefsbegincount=\referencecount
\immediate\write\referencewrite{\noexpand\refitem{#1.}}%
\begingroup\obeyendofline\rw@start}
\newcount\figurecount     \figurecount=0
\newif\iffigureopen       \newwrite\figurewrite
\def\figch@ck{\chardef\rw@write=\figurewrite \iffigureopen\else
   \immediate\openout\figurewrite=figures.texauxil
   \figureopentrue\fi}
\def\FIGNUM#1{\space@ver{}\figch@ck \firstreflinetrue%
 \global\advance\figurecount by 1 \xdef#1{\the\figurecount}}
\def\FIG#1{\FIGNUM#1
   \immediate\write\figurewrite{\noexpand\refitem{#1.}}%
   \begingroup\obeyendofline\rw@start}
\def\fig{\FIGNUM\? fig.~\?%
\immediate\write\figurewrite{\noexpand\refitem{\?.}}%
\begingroup\obeyendofline\rw@start}
\def\figure{\FIGNUM\? figure~\?
   \immediate\write\figurewrite{\noexpand\refitem{\?.}}%
   \begingroup\obeyendofline\rw@start}
\def\Fig{\FIGNUM\? Fig.~\?%
\immediate\write\figurewrite{\noexpand\refitem{\?.}}%
\begingroup\obeyendofline\rw@start}
\def\Figure{\FIGNUM\? Figure~\?%
\immediate\write\figurewrite{\noexpand\refitem{\?.}}%
\begingroup\obeyendofline\rw@start}
\newcount\tablecount     \tablecount=0
\newif\iftableopen       \newwrite\tablewrite
\def\tabch@ck{\chardef\rw@write=\tablewrite \iftableopen\else
   \immediate\openout\tablewrite=tables.texauxil
   \tableopentrue\fi}
\def\TABNUM#1{\space@ver{}\tabch@ck \firstreflinetrue%
 \global\advance\tablecount by 1 \xdef#1{\the\tablecount}}
\def\TABLE#1{\TABNUM#1
   \immediate\write\tablewrite{\noexpand\refitem{#1.}}%
   \begingroup\obeyendofline\rw@start}
\def\Table{\TABNUM\? Table~\?%
\immediate\write\tablewrite{\noexpand\refitem{\?.}}%
\begingroup\obeyendofline\rw@start}
%

%
%
%
\def\masterreset{\global\pagenumber=1 \global\chapternumber=0
   \ifnum\the\equanumber<0\else \global\equanumber=0\fi
   \global\sectionnumber=0
   \global\referencecount=0 \global\figurecount=0 \global\tablecount=0 }
\def\FRONTPAGE{\ifvoid255\else\vfill\penalty-2000\fi
      \masterreset\global\frontpagetrue
      \global\lastf@@t=0 \global\footsymbolcount=0}

\def\paperstyle{\letterstylefalse\normalspace\papersize}
\def\letterstyle{\letterstyletrue\singlespace\lettersize}
\def\papersize{\hsize=6.5truein\vsize=9.1truein\hoffset=-.3truein
               \voffset=-.4truein\skip\footins=\bigskipamount}
\def\lettersize{\hsize=6.5truein\vsize=9.1truein\hoffset=-.3truein
    \voffset=.1truein\skip\footins=\smallskipamount \multiply
    \skip\footins by 3 }
\paperstyle   
%
%
\def\MEMO{\letterstyle\FRONTPAGE \letterfrontheadline={\hfil}
    \line{\quad\fourteenrm CERN MEMORANDUM\hfil\twelverm\the\date\quad}
    \medskip \memod@f}

\def\memit@m#1{\smallskip \hangafter=0 \hangindent=1in
      \Textindent{\caps #1}}
\def\memod@f{\xdef\mto{\memit@m{To:}}\xdef\from{\memit@m{From:}}%
     \xdef\topic{\memit@m{Topic:}}\xdef\subject{\memit@m{Subject:}}%
     \xdef\rule{\bigskip\hrule height 1pt\bigskip}}
\memod@f
\newskip\lettertopfil
\lettertopfil = 0pt plus 1.5in minus 0pt
\newskip\letterbottomfil
\letterbottomfil = 0pt plus 2.3in minus 0pt
\newskip\spskip \setbox0\hbox{\ } \spskip=-1\wd0
\def\addressee#1{\medskip\rightline{\the\date\hskip 30pt} \bigskip
   \vskip\lettertopfil
   \ialign to\hsize{\strut ##\hfil\tabskip 0pt plus \hsize \cr #1\crcr}
   \medskip\noindent\hskip\spskip}
\newskip\signatureskip       \signatureskip=40pt
\def\signed#1{\par \penalty 9000 \bigskip \dt@pfalse
  \everycr={\noalign{\ifdt@p\vskip\signatureskip\global\dt@pfalse\fi}}
  \setbox0=\vbox{\singlespace \halign{\tabskip 0pt \strut ##\hfil\cr
   \noalign{\global\dt@ptrue}#1\crcr}}
  \line{\hskip 0.5\hsize minus 0.5\hsize \box0\hfil} \medskip }

\def\endletter{\ifnum\pagenumber=1 \vskip\letterbottomfil\supereject
\else \vfil\supereject \fi}
\newbox\letterb@x
\def\lettertext{\par\unvcopy\letterb@x\par}
\def\multiletter{\setbox\letterb@x=\vbox\bgroup
      \everypar{\vrule height 1\baselineskip depth 0pt width 0pt }
      \singlespace \topskip=\baselineskip }
\def\letterend{\par\egroup}
%
%
%
\newskip\frontpageskip
\newtoks\pubtype
\newtoks\Pubnum
\newtoks\pubnum
\newtoks\pubnu
\newtoks\pubn
\newif\ifp@bblock  \p@bblocktrue
\def\PH@SR@V{\doubl@true \baselineskip=24.1pt plus 0.2pt minus 0.1pt
             \parskip= 3pt plus 2pt minus 1pt }
\def\PHYSREV{\paperstyle\PhysRevtrue\PH@SR@V}
\def\titlepage{\FRONTPAGE\paperstyle\ifPhysRev\PH@SR@V\fi
   \ifp@bblock\p@bblock\fi}
\def\nopubblock{\p@bblockfalse}
\def\endpage{\vfil\break}
\frontpageskip=1\medskipamount plus .5fil
\pubtype={\tensl Preliminary Version}
\Pubnum={$\rm CERN-TH.\the\pubnum $}
\pubnum={0000}
\def\p@bblock{\begingroup \tabskip=\hsize minus \hsize
   \baselineskip=1.5\ht\strutbox \topspace-2\baselineskip
   \halign to\hsize{\strut ##\hfil\tabskip=0pt\crcr
   \the \pubn\cr
   \the \Pubnum\cr
   \the \pubnu\cr
   \the \date\cr}\endgroup}
\def\title#1{\vskip\frontpageskip \titlestyle{#1} \vskip\headskip }
\def\author#1{\vskip\frontpageskip\titlestyle{\twelvecp #1}\nobreak}

\def\address#1{\par\kern 5pt\titlestyle{\twelvepoint\it #1}}
\def\andaddress{\par\kern 5pt \centerline{\sl and} \address}

\def\abstract{\vskip\frontpageskip\centerline{\fourteenrm ABSTRACT}
              \vskip\headskip }

%
%
%

\def\\{\relax\ifmmode\backslash\else$\backslash$\fi}
\def\globaleqnumbers{\relax\ifnum\the\equanumber<0%
\else\global\equanumber=-1\fi}
\def\nextline{\unskip\nobreak\hskip\parfillskip\break}

\def\journal#1&#2(#3){\unskip, \sl #1~\bf #2 \rm (19#3) }

\def\topspace{\hrule height 0pt depth 0pt \vskip}

\let\int=\intop         
\def\prop{\mathrel{{\mathchoice{\pr@p\scriptstyle}{\pr@p\scriptstyle}{
                \pr@p\scriptscriptstyle}{\pr@p\scriptscriptstyle} }}}
\def\pr@p#1{\setbox0=\hbox{$\cal #1 \char'103$}
   \hbox{$\cal #1 \char'117$\kern-.4\wd0\box0}}
\def\lsim{\mathrel{\mathpalette\@versim<}}
\def\gsim{\mathrel{\mathpalette\@versim>}}
\def\@versim#1#2{\lower0.2ex\vbox{\baselineskip\z@skip\lineskip\z@skip
  \lineskiplimit\z@\ialign{$\m@th#1\hfil##\hfil$\crcr#2\crcr\sim\crcr}}}
\def\leftrightarrowfill{$\m@th \mathord- \mkern-6mu
        \cleaders\hbox{$\mkern-2mu \mathord- \mkern-2mu$}\hfil
        \mkern-6mu \mathord\leftrightarrow$}
\def\lrover#1{\vbox{\ialign{##\crcr
        \leftrightarrowfill\crcr\noalign{\kern-1pt\nointerlineskip}
        $\hfil\displaystyle{#1}\hfil$\crcr}}}
%
%
%
\let\sec@nt=\sec
\def\sec{\relax\ifmmode\let\n@xt=\sec@nt\else\let\n@xt\section\fi\n@xt}
\def\obsolete#1{\message{Macro \string #1 is obsolete.}}
\def\firstsec#1{\obsolete\firstsec \section{#1}}
\def\firstsubsec#1{\obsolete\firstsubsec \subsection{#1}}
\def\thispage#1{\obsolete\thispage \global\pagenumber=#1\frontpagefalse}
\def\thischapter#1{\obsolete\thischapter \global\chapternumber=#1}
\def\nextequation#1{\obsolete\nextequation \global\equanumber=#1
   \ifnum\the\equanumber>0 \global\advance\equanumber by 1 \fi}
\def\BOXITEM{\afterassigment\B@XITEM\setbox0=}
\def\B@XITEM{\par\hangindent\wd0 \noindent\box0 }
%

%
%

%
%

%
%

%

%

%

%

%

%
%
%
\def\boxit#1{\vbox{\hrule\hbox{\vrule\kern3pt\vbox{\kern3pt#1\kern3pt}
\kern3pt\vrule}\hrule}}
%
%
%
\catcode`@=12 
\message{ by V.K./U.B.}
\everyjob{\input imyphyx }
%
%
%
%
%
%
%
%
%
%
%
%
%
\catcode`@=11

\font\seventeencp=cmcsc10 scaled\magstep3
\def\SIZE{\hsize=6.6truein\vsize=9.1truein}
\def\OFFSET{\voffset=1.2truein\hoffset=.8truein}
\def\papersize{\SIZE\OFFSET\skip\footins=\bigskipamount
\normaldisplayskip= 30pt plus 5pt minus 10pt}
\Pubnum={\rm CERN$-$TH.\the\pubnum }
\def\title#1{\vskip\frontpageskip\vskip .50truein
     \titlestyle{\seventeencp #1} \vskip\headskip\vskip\frontpageskip
     \vskip .2truein}
\def\author#1{\vskip .27truein\titlestyle{#1}\nobreak}

\def\p@bblock{\begingroup \tabskip=\hsize minus \hsize
   \baselineskip=1.5\ht\strutbox \topspace-2\baselineskip
   \halign to\hsize{\strut ##\hfil\tabskip=0pt\crcr
   \the \Pubnum\cr}\endgroup}
\def\makefootline{\iffrontpage\vskip .27truein\line{\the\footline}
                 \vskip -.1truein\line{\the\date\hfil}
              \else\line{\the\footline}\fi}
\paperfootline={\iffrontpage \the\Pubnum\hfil\else\hfil\fi}
\paperheadline={\iffrontpage\hfil
                \else\twelverm\hss $-$\ \folio\ $-$\hss\fi}
\newif\ifmref  
\newif\iffref  
\def\xrefsend{\xrefmark{\count255=\referencecount
\advance\count255 by-\lastrefsbegincount
\ifcase\count255 \number\referencecount
\or \number\lastrefsbegincount,\number\referencecount
\else \number\lastrefsbegincount-\number\referencecount \fi}}
\def\xrefsdub{\xrefmark{\count255=\referencecount
\advance\count255 by-\lastrefsbegincount
\ifcase\count255 \number\referencecount
\or \number\lastrefsbegincount,\number\referencecount
\else \number\lastrefsbegincount,\number\referencecount \fi}}
\def\xREFNUM#1{\space@ver{}\refch@ck\firstreflinetrue%
\global\advance\referencecount by 1
\xdef#1{\xrefend}}
\def\xrefend{\xrefmark{\number\referencecount}}
\def\xrefmark#1{[{#1}]}
\def\xRef#1{\xREFNUM#1\immediate\write\referencewrite%
{\noexpand\refitem{#1}}\begingroup\obeyendofline\rw@start}%
\def\xREFS#1{\xREFNUM#1\global\lastrefsbegincount=\referencecount%
\immediate\write\referencewrite{\noexpand\refitem{#1}}%
\begingroup\obeyendofline\rw@start}
\def\rrr#1#2{\relax\ifmref{\iffref\xREFS#1{#2}%
\else\xRef#1{#2}\fi}\else\xRef#1{#2}\xrefend\fi}
\referencecount=0
%
\space@ver{}\refch@ck\firstreflinetrue%
\immediate\write\referencewrite{}%
\begingroup\obeyendofline\rw@start{}%
\def\plb#1({Phys.\ Lett.\ $\underline  {#1B}$\ (}
\def\nup#1({Nucl.\ Phys.\ $\underline {B#1}$\ (}
\def\plt#1({Phys.\ Lett.\ $\underline  {B#1}$\ (}
\def\cmp#1({Comm.\ Math.\ Phys.\ $\underline  {#1}$\ (}
\def\prp#1({Phys.\ Rep.\ $\underline  {#1}$\ (}
\def\prl#1({Phys.\ Rev.\ Lett.\ $\underline  {#1}$\ (}
\def\prv#1({Phys.\ Rev. $\underline  {D#1}$\ (}
\def\und#1({            $\underline  {#1}$\ (}
\message{ by W.L.}
\everyjob{\input offset }
\catcode`@=12

\let\it=\sl

\def\OFFSET{\hoffset=6.pt\voffset=40.pt}
\def\SIZE{\hsize=420.pt\vsize=620.pt}
\OFFSET
\def\PLANCK{\rrr\PLANCK{L.E. Ib\'a\~nez, \plb126 (1983) 196;
\nextline J.E. Bjorkman and D.R.T. Jones, \nup259 (1985) 533.}}

\def\ZNZM{\rrr\ZNZM{A. Font, L.E. Ib\'a\~nez and F. Quevedo,
\plb217 (1989) 272.}}

\def\IN{\rrr\IN{L.E. Ib\'a\~nez and H.P. Nilles, \plb169 (1986) 354.}}

\def\EINHJON{\rrr\EINHJON{M. Einhorn and D.R.T. Jones, \nup196 (1982)
                      475.}}

\def\ABK{\rrr\ABK{I. Antoniadis, C. Bachas and C. Kounnas,
        \nup289 (1987) 87.}}

\def\KAPLU{\rrr\KAPLU{V. Kaplunovsky, \nup307 (1988) 145.}}

\def\AMALDI{\rrr\AMALDI{
J. Ellis, S. Kelley and D.V. Nanopoulos, \plb249 (1990) 441;
\plb260 (1991) 131;
P. Langacker, {\it ``Precision tests of the standard model",}
Pennsylvania preprint UPR-0435T, (1990);
U. Amaldi, W. de Boer and H. F\"urstenau, \plt260 (1991) 447;
P. Langacker and M. Luo, Pennsylvania preprint UPR-0466T, (1991);
R. Roberts and G.G. Ross, talk presented by G.G. Ross at 1991 Joint
International Lepton-Photon Symposium and EPS Conference and to
be published.
}}

\def\KAC{\rrr\KAC{ A. Font, L.E. Ib\`a\~nez and F. Quevedo,
\nup345 (1990) 389;
J. Ellis, J. L\'opez and D.V. Nanopoulos,  \plb245 (1990) 375.}}

\def\DG{\rrr\DG{S. Dimopoulos, S. Raby and F. Wilczek, Phys. Rev.
D24 (1981) 1681;\nextline
                 L.E. Ib\'a\~nez and G.G. Ross, \plb105 (1981) 439;
\nextline S. Dimopoulos and H. Georgi, \nup193 (1981) 375.}}

\def\IMNQ{\rrr\IMNQ{L.E. Ib\'a\~nez, H.P. Nilles and F. Quevedo,
\plt187 (1987) 25; L.E. Ib\'a\~nez, J. Mas, H.P. Nilles and
F. Quevedo, \nup301 (1988) 157; A. Font, L.E. Ib\'a\~nez,
F. Quevedo and A. Sierra, \nup331 (1990) 421.}}

\def\SCHELL{\rrr\SCHELL{A.N. Schellekens, \plt237 (1990) 363.}}

\def\GQW{\rrr\GQW{H. Georgi, H.R. Quinn and S. Weinberg, Phys. Rev.
Lett. ${\underline{33}}$ (1974) 451.}}

\def\GINS{\rrr\GINS{P. Ginsparg, \plt197 (1987) 139.}}

\def\ELLISETAL{\rrr\ELLISETAL{I. Antoniadis, J. Ellis, R. Lacaze
and D.V. Nanopoulos, {\it ``String Threshold Corrections and
                Flipped $SU(5)$'',} preprint CERN-TH.6136/91 (1991);
    S. Kalara, J.L. Lopez and D.V. Nanopoulos, {\it``Threshold
   Corrections and Modular Invariance in Free Fermionic Strings'',}
      preprint CERN-TH-6168/91 (1991).}}

\def\LLS{\rrr\LLS{W. Lerche, D. L\"ust and A.N. Schellekens, \nup287
         (1987) 477.}}

\def\CREMMER{\rrr\CREMMER{E. Cremmer, S. Ferrara, L. Girardello and
          A. Van Proeyen, \nup212 (1983) 413.}}

\def\DFKZ{\rrr\DFKZ{J.P. Derendinger, S. Ferrara, C. Kounnas and F. Zwirner,
         {\it ``On loop corrections to string effective field theories:
         field-dependent gauge couplings and sigma-model anomalies'',}
        preprint CERN-TH.6004/91, LPTENS 91-4 (revised version) (1991).}}

\def\LOUIS{\rrr\LOUIS{J. Louis, {\it
         ``Non-harmonic gauge coupling constants in supersymmetry
         and superstring theory'',} preprint SLAC-PUB-5527 (1991);
         V. Kaplunovsky and J. Louis, as quoted in J. Louis,
         SLAC-PUB-5527 (1991).}}

\def\DIN{\rrr\DIN{J.P. Derendinger, L.E. Ib\'a\~nez and H.P Nilles,
        \nup267 (1986) 365.}}

\def\DHVW{\rrr\DHVW{L. Dixon, J. Harvey, C.~Vafa and E.~Witten,
         \nup261 (1985) 651;
        \nup274 (1986) 285.}}

\def\DKLB{\rrr\DKLB{L. Dixon, V. Kaplunovsky and J. Louis,
         \nup355 (1991) 649.}}

\def\DKLA{\rrr\DKLA{L. Dixon, V. Kaplunovsky and J. Louis, \nup329 (1990)
            27.}}

\def\FKLZ{\rrr\FKLZ{S. Ferrara, C. Kounnas, D. L\"ust and F. Zwirner,
           {\it ``Duality Invariant Partition Functions and
           Automorphic Superpotentials for (2,2) String
           Compactifications''}, preprint CERN-TH.6090/91 (1991),
           to appear in Nucl. Phys. B.}}

\def\ALOS{\rrr\ALOS{E. Alvarez and M.A.R. Osorio, \prv40 (1989) 1150.}}

\def\FILQ{\rrr\FILQ{A. Font, L.E. Ib\'a\~nez, D. L\"ust and F. Quevedo,
           \plt245 (1990) 401.}}

\def\MAGN{\rrr\MAGN{S. Ferrara, N. Magnoli, T.R. Taylor and
           G. Veneziano, \plt245 (1990) 409.}}

\def\CFILQ{\rrr\CFILQ{M. Cvetic, A. Font, L.E.
           Ib\'a\~nez, D. L\"ust and F. Quevedo, \nup361 (1991) 194.}}

\def\MOD{\rrr\MOD
        {R.~Dijkgraaf, E.~Verlinde and H.~Verlinde, \cmp115 (1988) 649;
        {\it ``On moduli spaces of conformal field theories with $c\geq
        1$'',} preprint THU-87/30;
           A. Shapere and F. Wilczek, \nup320 (1989) 669.}}

\def\DUAL{\rrr\DUAL{K. Kikkawa and M. Yamasaki, \plb149 (1984) 357;
           N. Sakai and I. Senda, Progr. Theor. Phys. 75 (1986) 692.}}

\def\FILQ{\rrr\FILQ{A. Font, L.E. Ib\'a\~nez, D. L\"ust and F. Quevedo,
           \plt245 (1990) 401.}}

\def\DUAGAU{\rrr\DUAGAU{S. Ferrara, N. Magnoli, T.R. Taylor and
           G. Veneziano, \plt245 (1990) 409; H.P. Nilles and M.
           Olechowski, \plt248 (1990) 268; P. Binetruy and M.K.
           Gaillard, \plt253 (1991) 119.}}
\def\TV{\rrr\TV{T.R. Taylor and G. Veneziano, \plt212 (1988) 147.}}

\def\CFILQ{\rrr\CFILQ{M. Cvetic, A. Font, L.E.
           Ib\'a\~nez, D. L\"ust and F. Quevedo, \nup361 (1991) 194.}}

\def\FIQ{\rrr\FIQ{A. Font, L.E. Ib\'a\~nez and F. Quevedo,
        \plt217 (1989) 272.}}

\def\FLST{\rrr\FLST{S. Ferrara,
         D. L\"ust, A. Shapere and S. Theisen, \plt225 (1989) 363.}}

\def\FLT{\rrr\FLT
{S. Ferrara, D. L\"ust and S. Theisen, \plt233 (1989) 147.}}

\def\CAOV{\rrr\CAOV{G. Lopes Cardoso and B. Ovrut, {\it ``A Green-Schwarz
            Mechanism for D=4, N=1 supergravity Anomalies,''}
            preprint UPR-0464T (1991); {\it ``Sigma Model Anomalies,
           Non-Harmonic gauge Couplings and String Theory,''}
           preprint UPR-0481T (1991).
            }}

\def\IBLU{\rrr\IBLU{L. Ib\'a\~nez and D. L\"ust, \plt267 (1991) 51.}}

\def\GAUGINO{\rrr\GAUGINO{J.P. Derendinger, L.E. Ib\'a\~nez and H.P. Nilles,
            \plb155 (1985) 65;
         M. Dine, R. Rohm, N. Seiberg and E. Witten, \plb156 (1985) 55.}}

\def\GHMR{\rrr\GHMR{D.J. Gross, J.A. Harvey, E. Martinec and R. Rohm,
         \prl54 (1985) 502; \nup256 (1985) 253; \nup267 (1986) 75.}}

\def\KLT{\rrr\KLT{H. Kawai, D.C. Lewellen and S.-H. H. Tye, \prl57 (1986)
        1832; (E) \und{58} (1987) 429; \nup288 (1987) 1.}}

\def\CHSW{\rrr\CHSW{P. Candelas, G. Horowitz, A. Strominger and
           E. Witten, \nup258 (1985) 46.}}

\def\ANT{\rrr\ANT{I. Antoniadis, K.S. Narain and T.R. Taylor,
        {\it ``Higher Genus String Corrections to Gauge Couplings'',}
         preprint NUB-3025 (1991).}}

\def\FERRA{\rrr\FERRA{S. Ferrara, preprint UCLA/90/TEP/20 (1990).}}

\def\GS{\rrr\GS{M.B. Green and J.H. Schwarz, \plb149 (1984) 117.}}

\def\WEIN{\rrr\WEIN{S. Weinberg, \plb91 (1980) 51.}}

\def\WITTEF{\rrr\WITTEF{E. Witten, \plb155 (1985) 151.}}

\def\OFFSET{\hoffset=12.pt\voffset=55.pt}
\def\SIZE{\hsize=420.pt\vsize=620.pt}

\catcode`@=12
\newtoks\Pubnumtwo
\newtoks\Pubnumthree
\catcode`@=11
\def\p@bblock{\begingroup\tabskip=\hsize minus\hsize
   \baselineskip=0.5\ht\strutbox\topspace-2\baselineskip
   \halign to \hsize{\strut ##\hfil\tabskip=0pt\crcr
   \the\Pubnum\cr  \the\Pubnumtwo\cr 
   \the\pubtype\cr}\endgroup}
\pubnum={6241/91}
\date{September 1991}
\pubtype={}
\titlepage
\vskip -.6truein
\title{
Gauge Coupling Running in Minimal $SU(3)\times SU(2)\times U(1)$
Superstring Unification   }
 \centerline{\bf Luis E. Ib\'a\~nez}
 \vskip .1truein
   \vskip 0.1truein
  \centerline{\bf Dieter L\"ust}
  \vskip 0.1truein
  \centerline{CERN, 1211 Geneva 23, Switzerland}
\vskip 0.1truein
\centerline{and}
\centerline{\bf Graham G. Ross}
\vskip 0.1truein
\centerline{Dep. Theoretical Physics, 1 Keble Rd., Oxford, England}
\abstract\noindent\nobreak
We study the evolution of the gauge coupling
constants in string unification schemes in which the light
spectrum below the compactification scale
is exactly that of the minimal supersymmetric
standard model. In the absence of string threshold corrections
the predicted values $\sin^2\theta _W=0.218$ and $\alpha _s=0.20$
are in gross conflict with experiment, but these  corrections
are generically important.
One can express the string threshold  corrections to $\sin^2\theta _W$
and $\alpha_s$ in terms of certain
$modular$ $weights$ of quark, lepton and
Higgs superfields as well as the $moduli$ of the string model.
We find that in order to get agreement with the experimental
measurements within the context of this $minimal$ scheme,
certain constraints on the $modular$ $weights$ of the quark, lepton
and Higgs superfields should be obeyed.
Our analysis indicates that this $minimal$ $string$ $unification$
scheme is a              rather constrained scenario.
\vskip 1.0cm
\endpage
\pagenumber=1
\sequentialequations

Unification of  coupling constants is a necessary phenomenon in string
theory.
Specifically, at tree level, the gauge
couplings of a gauge group $G_a$ have  simple relations\GINS\
to the string coupling constant which is determined by the
vacuum expectation value of the dilaton field:
${1\over g_a^2}={k_a\over g_{\rm string}^2}$ where $k_a$ is the
level of the corresponding Kac-Moody algebra. At higher loop levels
this relation holds only at the typical string scale which is of the
order of the Planck mass $M_P$. Below this scale all couplings evolve
according to their renormalization group equations in the same way
as in standard $GUT$ theories as first discussed in \GQW.
This allows a comparison of the coupling constants with
the low energy data considering a specific string model.
In addition, thresholds effects due the massive string excitations
modify the above mentioned tree level relations.

Let us recall briefly the exact definition of the string mass scale.
It is given in the ${\overline{MS}}$ scheme by \KAPLU
$$M_{\rm string}^2
= {{2\ e^{(1-\gamma )}}\over {{\sqrt {27}}\pi {\alpha '}}}
\eqn \mstring $$
where $\gamma $ is the Euler constant and
$\alpha ' = 16\pi /g_{\rm string}^2M_{P}^2$. Numerically one finds \DFKZ
$$M_{\rm string}=0.7\times g_{\rm string}  \times 10^{18}~{\rm GeV}.
\eqn\stringsca
$$
(Note that this value differs from the one found in \KAPLU.)
This mass scale has to be compared with low energy data using the
field theory renormalization group equations and taking into account
also the model-dependent
stringy threshold corrections. A phenomenological very
promising model is the minimal supersymmetric standard model
with gauge group $G=SU(3)\times SU(2)\times U(1)$. The relevant evolution
of the electro-weak and strong coupling constants was considered some time
ago in \DG,\EINHJON. Recently this analysis was reconsidered \AMALDI\
taking into account the up-dated low energy data. The results
for $\sin^2\theta _W$ and    $\alpha _s$ are in very good agreement
with data for a value of the unification mass
 $M_X \simeq         10^{16}~{\rm GeV}$
and a susy threshold close to the weak scale.
On the other hand, as we show below, the large value for the
string unification scale $M_{string}$ leads to rather embarrassing
results for the couplings $sin^2\theta _W^0=0.218$ and
$\alpha _s^0=0.20$.
In this paper we discuss the question whether
one can make consistent the
unification scale of the minimal supersymmetric standard model
with the
relevant string unification scale $M_{\rm string}$.
(String unification and threshold effects within the flipped
$SU(5)$ model were considered in \ELLISETAL.)
At the first sight, this seems very unlikely since $M_{\rm string}$
is substantially
larger than the minimal susy model scale $M_X$. However, one
might hope that the effects of the string threshold contributions
could make
the separation of these two scales consistent. Although the
threshold effects are rather small in usual grand unified
models \WEIN\
it is not obvious that the same holds true in string unification
since we have to remember that above the  string
scale an infinite number of massive states
contribute to the threshold. This is obviously very different
from field theory unification scenarios.

The structure of the paper is the following.
First we will collect some formulas about one-loop gauge
coupling constants with special emphasis on string threshold
corrections and their relation to target space duality.
Then we will apply these formulas
to the case of the
unification of the three physical coupling constants $g_1,g_2,g_3$.
Our approach here will be mainly phenomenological.
We will consider a possible situation in which

\noindent a) the massless
particles with standard model gauge couplings are just those of the
minimal supersymmetric standard model

\noindent b) there is no partial
(field theoretical) unification scheme below the string scale.

\noindent
This is in principle the simplest string unification scheme that
one can think of and that is why we call it $minimal$ $string$
$unification$. Up to now no realistic string model with this
characteristics has been built but the model search done up to now
is extremely limited and by no means complete. We would like to answer
the question whether such a minimal scenario can be made consistent
with the measured values of the low energy coupling constants.

The one-loop running gauge coupling constant of a (simple) gauge
group $G_a$ is of the following form:
$${1\over g_a^2(\mu)}={k_a\over  g_{\rm string}^2}+{b_a\over 16\pi^2}
\log{M_a^2\over\mu^2}+\Delta_a.\eqn\rgea
$$
Here $b_a=-3C(G_a)+\sum_{R_a}h_{R_a}T(R_a)$
is the $N=1$ $\beta$-function coefficient
($h_{R_a}$ is the number of chiral matter fields in a representation
${\underline R}_a$).
$M_a$ is the renormalization
point below which the effective field theory running of the coupling
constant begins. (As we will discuss below, $M_a$ will
depend on the specific  model and also on the considered gauge group.)
$\Delta_a$ are the string threshold contributions \KAPLU\ which
arise due to the integration
over the infinite number of massive string
states, in particular momentum and winding states: $\Delta_a\propto\log
\det {\cal M}$, where ${\cal M}$ is the mass matrix of the heavy modes.

In the following we would like to give a brief description of how
one derives the expressions for the field-dependent stringy threshold
corrections and for the renormalization scale $M_a$
which, in general, is also field dependent.
Most directly, these quantities can be
obtained by world-sheet string computations
of string amplitudes involving external gauge fields and moduli
as done \DKLB,\ANT\ for
the case of (2,2) symmetric
orbifold compactifications \DHVW.
These computations are closely
related to the calculation \FKLZ\ of the target space free energies of
compactified strings.

A second very useful approach to obtain information about the form
of the string threshold corrections is the use of the target space
duality symmetries \DUAL\ present in many known string compactifications.
Here, the main idea is related to the observation \TV\ that in
string compactifications the scale $M_a$ below which the effective
field theory running of the gauge coupling constants starts becomes
a moduli dependent quantity,
$$M^2_a=(2R^2)^\alpha M^2_{\rm string}.\eqn\radius
$$
$R$ is a background parameter denoting
(in Planck units) the
overall radius of the compact six-dimensional space, and the power
$\alpha$ is
a model- and gauge group dependent parameter.
(In naive field theory compactifications one expects $\alpha=-1$.
However, as we will discuss in the following,
for orbifold compactifications $\alpha$ can also take different
values.) Thus the running gauge coupling constant eq.\rgea\
generically depends on the background radius.
To be specific consider orbifold type of compactifications. Here
the radius is related to the real part of a complex modulus field, $T=R^2
+iB$, ($B$ is an internal axion field) and the target space duality
group is given \MOD\
by the modular group $PSL(2,{\bf Z})$, acting on $T$ as
$T\rightarrow{aT-ib\over icT+d}$ ($a,b,c,d\in{\bf Z}$, $ad-bc=1$).
It follows that the effective  action involving the $T$-field
must be target space modular invariant and is given in terms of
modular functions \FLST.
Now, requiring \FILQ\ the invariance of $g_a^2(\mu)$ under target space
modular transformations enforces $\Delta_a$
to be a non-trivial $T$-dependent function.
Specifically, as discussed in \FILQ,\MAGN\ for the case $\alpha=-1$,
target space modular invariance, together with the requirement of having
no poles inside the fundamental region, implies
$$\Delta_a(T,\bar T)={\alpha b_a\over 16\pi^2}\log |\eta(T)|^4,\eqn
\dedekf
$$
where $\eta(T)$ is the Dedekind function. Notice that for large $T$ one
recovers the linear behavior found in ref.\IN\ .

The parameter $\alpha$ is intimately related to the $modular$
$weights$ of
the charged matter fields which transform non-trivially under the gauge
group $G_a$. To understand this, consider a standard supergravity,
Yang Mills field theory \CREMMER\ with massless
gauge singlet chiral moduli fields
$T_i$ and massless\foot{If some of the matter fields become massive due
to a trilinear superpotential one ends up with the same results about
the form of the threshold corrections \IBLU.}
charged chiral matter fields $\phi_i^{R_a}$ ($i=1,\dots ,h_{R_a}$).
The relevant part of the tree level
supergravity Lagrangian is specified by
the following K\"ahler potential at lowest order in $\phi_i^R$:
$$K(T_i,\phi_i^R)=K(T_i,\bar T_i)+K_{ij}^R(T_i,\bar T_i)
\phi_i^R\bar\phi_j^R.\eqn\kp
$$
In the following we assume
that the K\"ahler metric for the charged fields
is proportional to the K\"ahler metric of the moduli,
which was shown \DKLA\ to be
true for (2,2) Calabi-Yau string compactifications \CHSW, i.e.
$K_{ij}^R\propto\partial_{T_i}\partial_{\bar T_j}K(T_i,\bar T_i)$.
As discussed in \DFKZ,\LOUIS,\CAOV, at the one loop level
$\sigma$-model anomalies play a very important role for the
determination of the renormalized gauge coupling constant. Specifically,
one has to consider two types of triangle diagrams with two gauge bosons
and several moduli fields as external legs and massless gauginos and
charged (fermionic) matter
fields circulating inside the loop: First the coupling of the moduli
to the charged fields
contains a part
described by a composite K\"ahler connection, proportional to $K(T_i,
\bar T_i)$, which couples to gauginos as well as to chiral matter
fields $\phi_i^{R_a}$. It
reflects the (tree level) invariance of the theory under K\"ahler
transformations. Second, there is a coupling between the moduli
and the $\phi_i^{R_a}$'s
by the composite
curvature (holonomy) connection. It
originates from the non-canonical
kinetic energy $K_{ij}^R$
of the matter fields $\phi_i^R$ and shows the (tree level)
invariance of the theory under general coordinate transformations
on the complex moduli space. These two anomalous contributions
lead, via supersymmetry, to the following one-loop modification
of the gauge coupling constant \LOUIS,\DFKZ,\CAOV:
$$\eqalign{{1\over g_a^2}&={k_a\over g_{\rm string}^2}-{1\over 16\pi^2}
\biggl((C(G_a)-\sum_{R_a}h_{R_a}T(R_a)) K(T_i,\bar T_i)\cr&+
2\sum_{R_a}T(R_a)\log\det K_{ij}^R(T_i,\bar T_i)\biggr) .\cr}
\eqn\nl
$$

Now assume that the string theory is invariant under
target space duality transformations which are discrete
reparametrizations of the moduli.
(The simple $R\rightarrow 1/R$ duality symmetry in bosonic string
compactification was shown \ALOS\ to be unbroken
in each order of string perturbation.)
These transformations do not leave invariant the K\"ahler potential
$K(T_i,\bar T_i)$ and also $\log\det K_{ij}$. Thus
eq.\nl\ is not invariant under duality transformations.
It follows that the duality anomalies
must be cancelled by adding
new terms to the effective action.
Specifically, there are two ways to cancel these anomalies.
First \DFKZ,\CAOV,
one can perform a moduli dependent, but gauge group independent
redefinition of the dilaton/axion
field, the socalled $S$-field,
such that $S+\bar S$ transforms
non-trivially under duality transformations and cancels in this
way some part or all of the duality non-invariance of eq.\nl.
This field redefinition of the $S$-field is analogous to
Green-Schwarz mechanism \GS\ and
leads to a mixing between the moduli and the $S$-field in the
$S$-field K\"ahler potential.
Second, the duality anomaly can be cancelled
by adding to eq.\nl\ a term which describes the threshold contribution
due to the massive string states. (Only the specific knowledge about
the massive string spectrum can determine the exact coefficients
for the Green-Schwarz and threshold terms whose combined variation
cancels the total modular anomaly. However, as it will become clear
in the following, the coefficient of
Green-Schwarz term is irrelevant for the
determination of the unificaton mass scales.)
In analogy to the Dedekind function
the threshold contributions are given in
terms of automorphic functions of the target space duality group.
Specifically, as described in \FKLZ, for general (2,2) Calabi-Yau
compactifications there exist two types of automorphic functions:
the first one provides a duality
invariant completion of $K(T_i,\bar T_i)$, where the second one
is needed to cancel the duality anomaly coming from $\log\det K_{ij}$.
These two types of automorphic functions can be, at least formally,
constructed for all (2,2) Calabi-Yau compactifications \FKLZ.

In the following we restrict ourselves to symmetric (but not
necessarily (2,2) symmetric) ${\bf Z}_N$  \DHVW,\IMNQ\  and
${\bf Z}_N\times{\bf Z_M}$ orbifolds \ZNZM.
Every orbifold of this type has
three complex planes corresponding to three
two-dimensional subtori. For non-trivial examples each orbifold twist
$\delta_m=(\delta_1,\delta_2,\delta_3)$ acts either simultaneously
on two or all three planes.
For simplicity we consider only the overall modulus $T=R^2+iB$ where
the target space duality transformations are given by the modular
group $PSL(2,{\bf Z})$. The K\"ahler potential for this overall
modulus looks like \WITTEF
$$K(T,\bar T)=-3\log(T+\bar T).\eqn\kpt
$$
The K\"ahler metric of the matter fields has the following
generic form \DKLA:
$$K_{ij}^{R_a}=\delta_{ij}(T+\bar T)^{n_{R_a}}.\eqn\kinen
$$
Target space  modular invariance then implies that the matter fields
transform under $PSL(2,{\bf Z})$ as
$$\phi_i^{R_a}\rightarrow\phi_i^{R_a}(icT+d)^{n_{R_a}}.\eqn\modwei
$$
Thus we identify the integers $n_{R_a}$ as the modular weights
of $\phi_i^{R_a}$. (For the ${\bf Z}_3$ orbifold see \FLT.)
Specifically, for symmetric orbifold compactifications
there are three different types of matter fields:

\noindent a) Untwisted matter with $n_R=-1$.

\noindent b) Twisted matter fields associated with an orbifold twist
$\delta_m$. Here $n_R=-2$ if $\delta_m$ acts on all three planes.
$n_R=-1$ if the twist acts only on two of the three planes.
Thus the latter
kind of twisted fields behave exactly like untwisted fields under
modular transformations.

\noindent c) Twisted moduli with $n_R=-3$ or $n_R=-2$
if the corresponding twist acts
on all three planes or only on two planes respectively.

\noindent Then, using eq.\nl,
the one-loop contribution to the gauge coupling
constant due to the anomalous triangle
diagrams with massless charged fields
has the following form \LOUIS,\DFKZ:
$$\eqalign{{1\over g_a^2}&={k_a\over g_{\rm string}^2}+{1\over 16\pi^2}
b_a'\log(T+\bar T),\cr
b_a'=
3C(G_a)&-\sum_{R_a}h_{R_a}T(R_a)(3+2n_{R_a})=-b_a-2\sum_{R_a}h_{R_a}
T(R_a)(1+n_{R_a}).\cr}
\eqn\nla
$$
As discussed already, the modular anomaly of this contribution
to ${1\over g_a^2}$ from the massless fields can be cancelled by
by a universal Green-Schwarz term plus the threshold contribution
from the massive orbifold excitations. The orbifold threshold contribution
takes the following form (up to a small $T$-independent term \KAPLU):
$$\Delta_a(T,\bar T)={1\over 16\pi^2}(b_a'-k_a
b_{GS})\log |\eta(T)|^4.\eqn
\threorbi
$$
Here $b_{GS}$ is the universal coefficient of the Green-Schwarz term.
Without going into any detail let us just state the main result
concerning the coefficient $b_a'-k_ab_{GS}$ \DKLB,\ANT.
The threshold contribution of the massive fields, i.e. $b_a'-k_ab_{GS}$,
is non-vanishing if at least one
of the three complex planes is not rotated
by some of the orbifold twist $\delta_m$. Then, within this sector,
the massive spectrum
with $T$-dependent masses
is $N=2$ space-time supersymmetric and $b_a'-k_ab_{GS}$ is proportional
to the $N=2$ $\beta$-function coefficient. In this
case $b_a'-k_ab_{GS}$ is in general non-zero for all gauge groups
including the unbroken $E_8$ in the hidden sector.
On the other hand, sectors corresponding to planes which are
rotated by all twists $\delta_i$ lead to a massive $T$-dependent spectrum
with $N=4$ space-time supersymmetry and therefore do not contribute to
the threshold corrections.

Let us insert the threshold contribution eq.\threorbi\
into the one-loop
running
coupling constant eq.\rgea:
$$\eqalign{
{1\over g_a^2(\mu)}&={k_a\over  g_{\rm string}^2}+{b_a\over 16\pi^2}
\log{M_a^2\over\mu^2}+
{1\over 16\pi^2}(b_a'-k_ab_{GS})\log |\eta(T)|^4,\cr
M_a^2&=(T_R)^\alpha
M^2_{\rm string}, \qquad \alpha={b_a'-k_ab_{GS}\over b_a},
\cr}
\eqn\rgeb
$$
where $T_R=T+\bar T=2R^2$.
This expression is explicitly target space modular invariant.
Here we have absorbed the
piece from the massless fields in eq.\nla\ which is not cancelled
by the Green-Schwarz term into the definition of the renormalization
point $M_a$ (the remainder is absorbed into ${1\over g_{\rm string}^2}$
\DFKZ)
since it is a field theoretical,
infrared effect and does not originate from
the heavy string modes.

Now we are finally ready to discuss the unification of the gauge
coupling constants. The unification mass scale $M_X$ where two gauge
group coupling constants become equal, i.e. ${1\over k_ag_a^2(M_X)}=
{1\over k_bg_b^2(M_X)}$, becomes using eq.\rgeb
$${M_X\over M_{\rm string}}=
\lbrack
T_R|\eta(T)|^4\rbrack^{{b_a'k_b-b_b'k_a
\over 2(b_ak_b-b_bk_a)}}
.\eqn\unific
$$
Note that since we are interested only in the difference of two
gauge couplings, the universal Green-Schwarz term is irrelevant for $M_X$.
Since the moduli-dependent function
$(T+\bar T)|\eta(T)|^4<1$ for all $T$
it follows that $M_X/M_{\rm string}$ is smaller (bigger) than one if
${b_a'k_b-b_b'k_a\over b_ak_b-b_bk_a}$ is bigger (smaller) than zero.
Comparing with the definition of $b_a'$ in
eq.\nla\ one recognizes that for $k_a=k_b$
twisted states with $n_R<-1$ are
necessarily required to have $M_X<M_{\rm string}$.

Let us now briefly discuss three known (2,2) orbifold examples ($k=1$).
First for the ${\bf Z}_3$
and ${\bf Z}_7$ orbifolds, each of the three planes is
simultaneously rotated by all twists. Thus $b_a'-b_{GS}=0$ for all
gauge groups. It trivially follows that the renormalization point
is given as $M_a=M_{\rm string}$. The radius independence of the
renormalization point is due to
the fact that the spectrum of the massive Kaluza-Klein and winding states
is $N=4$ supersymmetric and has therefore no effect in loop calculations.
The absence of threshold corrections (in other words, only a universal
gauge group independent piece contributes to the gauge coupling
constant at one loop) also trivially implies that the unification scale
$M_X$, for example the unification point of $E_8$
and $E_6$, is given by $M_{\rm string}$.

A second example, which is rather orthogonal to the previous case,
is the symmetric ${\bf Z}_2\times{\bf Z}_2$ orbifold. (Also many
four-dimensional heterotic strings
obtained by the fermionic \KLT,\ABK\ or by the covariant lattice \LLS\
construction fall into his category.) Here each of the ${\bf Z}_2$ twists
leaves invariant exactly one of the three orbifold planes. Then we obtain
that $b_{GS}=0$, i.e. there is no one-loop $S$--$T$ mixing in the
K\"ahler potential of this model. Furthermore, according to our general
rules all twisted matter fields (${\underline{27}}$ of $E_6$) have
modular weight $n_{27}=-1$ and behave like untwisted fields. It follows
that $b_a'=-b_a$ ($a=E_8,E_6$). Thus we obtain $\alpha=-1$ and the
radius dependence of the renormalization point agrees with the naive
field theoretical expectation: $M_a=M_{\rm string}T_R^{-1/2}$.
The unification scale of $E_8$ and $E_6$ is given by
$M_X=M_{\rm string}/(T_R^{1/2}|\eta(T)|^2)$ and is therefore
larger than the string scale for all values of the radius.

Now let us apply the above discussion to the case of the unification
of the gauge coupling constants within the minimal string unification.
We will make use of the threshold
formulae of eqs.\rgeb,\unific\
although they were originally derived for
a general class of abelian ${\bf Z}_N$ and
${\bf Z}_N\times {\bf Z}_M$  $(2,2)$
orbifolds. In fact the gauge groups in these cases is always
$E_6\times E_8$ and not anything looking like the standard model group.
However we would like to argue for the validity of these formulae
in the presence of Wilson lines and for $(0,2)$ type of gauge
embeddings because the structure of the untwisted moduli is
exactly the same as in the corresponding $(2,2)$ orbifold.
We will again
only consider the string threshold effects dependence on the
overall modulus $T$. Then the $T$-field K\"ahler potential and the
K\"ahler metric of the matter fields are given by eqs.\kpt\ and
\kinen\ respectively, and the low energy contribution to the gauge
coupling constants is still described by eq.\nla.
Thus, using the requirement of target space modular invariance,
the threshold formulae \rgeb,\unific\
remain valid for generic symmetric orbifolds, and not
only for their standard embeddings. (For example one can check that
for ${\bf Z}_3$ (0,2) orbifolds
the $b'$ coefficients of all gauge groups again
exactly  agree.)
These type of models may in
general yield strings with the gauge group of the standard model
and appropriate matter fields as discussed e.g. in \IMNQ.
In reality the threshold effects will depend not only on the
untwisted moduli but on other marginal deformations like the
twisted moduli and even on extra charged scalars with flat
potentials present in specific models. We believe that considering
just the dependence on the overall (volume) modulus gives us an idea
of the size and effects of the string threshold.
Finally, as discussed above, we would expect to find similar
results in more general (Calabi Yau) four dimensional strings
in which the threshold effects have not been explicitly computed.
The low energy anomaly arguments should be valid for an arbitrary
string and similar formulae to those below should be found for those
more general cases with the obvious replacements due to the
different duality groups involved.

Let us first consider the joining of the $SU(2)$ and $SU(3)$
gauge coupling constants $g_2$ and $g_3$ at a field theory
unification scale $M_X$. If such a unification takes place
eq.\unific\
leads to the result ($k_2=k_3=1$)
$$
M_X^2\ =\ M_{\rm string}^2
\ (T_R\ |\eta (T)|^4)^{{b_2'-b_3'}\over {b_2-b_3}}
\eqn \unif $$
for the unification scale of the $g_2$ and $g_3$ coupling
constants. Recalling that one always has $T_R|\eta (T)|^4\leq 1$,
one concludes that $M_X$ may be smaller or bigger than $M_{\rm string}$
depending on the relative sign of $(b_2'-b_3')$ versus
$(b_2-b_3)$. In general, the definition of $b_i'$ in eq.\nla\
shows that relative
sign depends on the modular weights of the matter fields.
In some cases (e.g. when all matter fields have modular weight
$n=-1$) one has $b_i'=-b_i$ and $M_X$ is necessarily bigger than
$M_{\rm string}$.
In these cases one cannot accommodate the difference between
$M_X$ and $M_{\rm string}$ we discussed above and the $minimal$ $string$
$unification$ scheme is simply not viable. This is the case of
any model based on the ${\bf Z}_2\times
{\bf Z}_2$ orbifold (or equivalent
models constructed with free world-sheet fermions) since all
matter fields have modular weight one.

Let us now be a bit more quantitative and try to answer the
following question: what are the values for modular weights $n_{\beta}$
of quarks, leptons and Higgs as well as the corresponding
values of $T_R$ which would
allow for ${\sin^2\theta _W}$ and $\alpha_s$ values in
reasonable agreement with data? Making use of equation \unific\
one gets for the value of the electroweak angle $\theta _W$
after some standard algebra
$$
{\sin^2\theta _W}(\mu) \ =\
{{k_2}\over {k_1+k_2}} - {{k_1}\over {k_1+k_2}}
{{\alpha_e(\mu)}\over {4\pi}}\biggl(  A\ \log({{M_{\rm string}^2}
\over {\mu ^2}}) - \ A'\
\log(T_R|\eta (T)|^4)\biggr)
\eqn \sint$$
where $A$ is given by
$$
A\ \equiv \ {{k_2}\over {k_1}}b_1-b_2
\eqn \AAA $$
and $A'$ has the same expression after replacing
$b_i\rightarrow   b_i'$.
The standard grand unification values of the Kac-Moody levels
correspond to the
choice $k_2=k_3=1$ and $k_1=5/3$. Finally, $\alpha _e$ is
the fine structure constant evaluated at a low energy scale
$\mu $ (e.g. $\mu =M_Z$). In an analogous way one can compute
the low energy value of the strong interactions fine
structure constant $\alpha _s$
$${1\over {\alpha _s(\mu )}}\ =\ {{k_3}\over {(k_1+k_2)}}
\biggl({1\over {\alpha_e(\mu)}}\ -\ {{1}\over {4\pi }}\ B\
\log({{M_{\rm string}^2}\over {\mu ^2}})\    - \
{{1}\over {4\pi }}\ B' \
\log(T_R|\eta (T)|^4)\biggr)   \eqn \alfs $$
where
$$
B\ \equiv \ b_1+b_2-{{(k_1+k_2)}\over {k_3}}b_3
\eqn \BBB $$
and $B'$ has the same expression after replacing
$b_i\rightarrow b_i'$. Let us now define
$$
\delta A\ \equiv \ A'\ +\ A\ =\
-2\ \sum_{\beta }(n_{\beta }+1)( {{k_2}\over k_1}\ {Y^2(\beta )}
    -                     T_2(\beta ) )
\eqn \dela $$
where the sum runs over all the matter fields and $n_{\beta }$
are the corresponding modular weights. $Y(\beta )$ is the
hypercharge         of each field and $T_2(\beta )$ the
corresponding $SU(2)$ quadratic Casimir ($T_2=1/2$ for a doublet).
Analogously let us     define
$$
\delta B \ =\ B\ + \ B'\ =\
-2\ \sum_{\beta }(n_{\beta }+1)\ ({Y^2(\beta )}+T_2(\beta )-
{{(k_1+k_2)}\over {k_3}}\ T_3(\beta )).
\eqn \delb $$
We can now write equations \sint\ and \alfs\ as follows ($k_2=
k_3=1$, $k_1=5/3$)
$${\sin^2\theta _W}\ =\ {3\over 8}\ -\
{{5{\alpha _e}}\over {32\pi}}\ A\ \log({{M_T^2}\over {\mu ^2}})\ -\
{{5{\alpha _e}}\over {32\pi}}\ \delta A\ \log(T_R|\eta (T)|^4),
\eqn \delaa $$
$$
{1\over {\alpha _s(\mu )}}\ =\ {3\over {8\alpha _e}}\ -\
{3B\over {32\pi}}\ \log({{M_T^2}\over {\mu ^2}})\ -\
{{3\delta B}\over {32\pi }}\ \log(T_R|\eta (T)|^4)
\eqn \delbb $$
where
$$M_T^2 \ \equiv \ {{M_{\rm string}^2}
\over {T_R|\eta (T)|^4}}.\eqn \mtt  $$

All the model dependence (through the modular weights) is
contained in $\delta A, \delta B$ . Denoting by
$n^i_{\beta }$ the modular weight of the $i-th$ generation
field of type $\beta =Q,U,D,L,E$ one can explicitly evaluate
that dependence and find
$$\delta A\ =\ {2\over 5}\sum_{i=1}^{N_{gen}}\ (
7n^i_Q+n^i_L
-4n^i_U-n^i_D-3n^i_E
)\ +\ {2\over 5}\ (2+n_H+n_{\bar H}),\eqn \delaaa $$
$$\delta B\ =\ 2\sum_{i=1}^{N_{gen}}\ (
n^i_Q+n^i_D-
n^i_L-n^i_E
)\ -
\  2\ (2+n_H+n_{\bar H}) \eqn \delbbb $$
where $N_{gen}$ is the number of generations and $n_H, n_{\bar H}$
are the modular weights of the Higgs fields.

It is easy to see from equations \sint\ and \alfs\ that both
$\delta A$ and $\delta B$ have to be positive in order to
have any chance to obtain the correct values for $\sin^2\theta _W$
and $\alpha _s$.
In the minimal supersymmetric standard model one has
$b_3=-3, b_2=m_H$ and $b_1=10+m_H$, where $m_H$ is the
number of pairs of Higgs doublets ($m_H=1$ in the minimal case).
For the standard unification $k_i$ values                 one then
finds $A=6-{2\over 5}m_H=28/5$ and
$B=18+2m_H=20$. As already explained, the requirement
$M_X<M_{\rm string}$ implies
$A'/A>0$ and also $B'/B>0$ (remember
$\log(T_R|\eta (T) |^4)$ is negative). Then one has the
conditions
$$ \delta A\ >    \ A\ =\ {{28}\over 5} , \eqn \consa $$
$$ \delta B\ >    \ B\ =\ 20   \eqn \consb $$
in the minimal model. Notice that these conditions are violated
explicitly in the ${\bf Z}_2\times {\bf Z}_2$ orbifolds (in which case
$\delta A=\delta B=0$) and also in ${\bf Z}_3$ and
${\bf Z}_7$. In the latter
cases one has $\delta A=A$ and $\delta B=B$ since $A'=B'=0$.
(Parenthetically,
these latter equations can be used combined with eqs.\delaaa\ and
\delbbb\  in order to get constraints on the number  of
$SU(2)$-doublets and $SU(3)$-triplets coming from untwisted, twisted
and twisted moduli sectors in specific ${\bf Z}_3$ and
${\bf Z}_7$ $(0,2)$ orbifolds).

Another point to remark is that if there
are $SU(5)$-type boundary conditions for the matter kinetic terms
(and, hence, for the modular weights) one has $n_U=n_Q=n_E$ and
$n_D=n_E$. In this case $\delta A ={2\over 5}(2+n_H+n_{\bar H})$
and $\delta B=-2(2+n_H+n_{\bar H})=-5\delta A$ and both quantities
cannot be simultaneously positive. Thus $SU(5)$-like boundary
conditions for the modular weights cannot accommodate the
values of the measured low energy couplings in the context of
minimal string unification. On the other hand there is no reason
why those boundary conditions should hold since we are assuming
that the gauge group is $SU(3)\times SU(2)\times U(1)$ up
to the string scale. Furthermore, other unification
schemes e.g. inside semisimple groups like $SU(4)\times SU(2)\times
SU(2)$ or $SU(3)\times SU(3)\times SU(3)$ do not lead to those
boundary conditions. All of these unification schemes are consistent
with $k_1=5/3$. Incidentally, let us recall that for the standard
values $k_2=k_3=1$ and $k_1=5/3$ the low energy symmetry is enlarged
to $SU(5)$ $only$ in the case we insist on the absence of massive
fractionally charged states \SCHELL. We do not insist on that, we just
assume the minimal low energy susy particle content but nothing specific
about the massive sector.

In principle, if the conditions \consa\  and \consb\   are met
there may exist a value of $T_R$ such that one can accommodate
the measured  low energy values of coupling constants.
In the absence of string threshold effects (i.e. for $\delta A=A$
and $\delta B=B$) one finds from equations \sint,\alfs\
$\sin^2\theta_W^0(M_Z)=0.218$ and $\alpha_s^0(M_Z)=0.205$. The effect of
non-vanishing threshold effects in the minimal scenario we are
discussing is displayed in figures 1 and 2. The first shows
the value of $\sin^2\theta _W(M_Z)$ as a function of
${\rm Re}T\equiv T_R/2$
for different values of
$\delta A$. A similar plot for $\alpha _s(M_Z)$ is shown in figure 2.
The shaded areas correspond to the experimental results.
The bounds in eqs.\consa\ and \consb\ are apparent in the figures.
One also observes that one can get results within the experimental
constraints for sufficiently large values of $\delta A,\delta B$ and
$T_R$.
In fact one can eliminate the explicit dependence on $T_R$ by
combining equations \delaa\ and \delbb. In this way one finds
a linear equation relating $\delta A$ and $\delta B$:
$$
(\delta B \ - \ B)\ =\ \gamma \ (\delta A \ - \ A),\eqn \lin $$
$$ \gamma \ =\ {5\over 3}{\alpha _e}\left( {{({1\over{ {\alpha _s}^0}}
-{1\over {\alpha _s(\mu )}})}\over {({sin^2\theta _W^0}-
sin^2\theta _W(\mu )}}\right).\eqn \bbet $$
If $\delta A/A=\delta B/B$ the string corrections may be entirely
contained in a change in the scale in the original field theoretical
analysis,
and all three coupling constants meet (at the one loop level)
at the same energy scale.
This corresponds to $\gamma=B/A=25/7$.
Allowing for
the experimental errors in $\alpha_s(\mu )$ and $\sin^2\theta _W(\mu )$
one more generally finds $2.2\leq \gamma \leq 4.0$.
One can then search for values for the modular weights $ n_{\beta }$ of
the standard model particles compatible with eqs.\consa,\consb,\lin.
Assuming generation independence for the $n_{\beta }$ as well
$-3\leq n_{\beta } \leq -1 $  one finds, interestingly enough,
a unique answer for the matter fields:
$$n_Q=n_D=-1\ \ ;\ \ n_U=-2 \ \ ;\ \  n_L=n_E=-3 \eqn \sol $$
and a constraint $n_H+n_{\bar H}=-5,-4$.
For $n_H+n_{\bar H}=-5$
one obtains
$\delta A=42/5$, $\delta B=30$,
and the three coupling constants meet at a scale $M_X\sim 2\times
10^{16}{\rm GeV}$ provided
that ${\rm Re}T$ is of order
${\rm Re}T\sim 16$ (see the two figures).
For $n_H+n_{\bar H}=-4$ one has
$\delta A=44/5$, $\delta B=28$. Now the three couplings only meet
approximately (within the experimental errors of $\sin^2\theta_W$
and $\alpha_s$) for similar values of ${\rm Re}T$.
Thus we see that, in principle, a situation with a
compactification scale below the string scale may work
within a minimal $SU(3)\times SU(2)\times U(1)$ string
provided, e.g., the above modular weights are possible.
Notice that (twisted) moduli fields are not
necessarily gauge singlets (e.g. the $SU(2)$ doublets in the ${\bf Z}_4$
orbifold).
 (Allowing for
    non-standard modular weights, i.e. $n<-3$,
the minimal unification scenario would     be possible for smaller
values of ${\rm Re}T$, and in particular for ${\rm Re}T\sim 1$.)

We have just shown that the $minimal$ $string$ $unification$ scenario
is in principle compatible with the measured low energy coupling
constants
for i) sufficiently large ${\rm Re}T$ and
ii) restricted choices of standard particles modular weights.
The question now is wether these two conditions are easy to meet.
Concerning the first condition, we need to have an idea
of the non-perturbative string
dynamics which trigger the compactification process
and fixes the value of ${\rm Re}T$. In the context of duality-invariant
effective actions, recent analysis \FILQ,\MAGN,\CFILQ\ shows that
the preferred values of ${\rm Re}T$ are of order
one. This is expected since a duality invariant potential will
typically have its minima not very far away from the self-dual
point. Thus large values of ${\rm Re}T$ are not expected within this
philosophy. However a deeper understanding about non-perturbative
string effects is definitely needed to give a final answer to this
question. Concerning the second condition,
it would be interesting to investigate ${\bf Z}_N$ and ${\bf Z}_M\times
{\bf Z}_N$ orbifold models to see whether
there are choices of modular weights leading to the appropriate
$\delta A,\delta B$. It is certainly intriguing the degree of
uniquenes in the possible choices of modular weights  (eq.\sol )
leading to adequate results and this point deserves further study.
Notice, however that one may relax the condition of generation
independence which led to eq.\sol. In addition one can consider
the separate contribution of the three orbifold planes in terms
of the three untwisted moduli $T_i$.
In summary, we believe that the
analysis presented in this paper shows that
$minimal$ $string$ $unification$  is a possible but rather constrained
scenario.

If no string model with the characteristics of the above
minimal unification scenario  is found,   it may still be possible to
explain the success of the unification of
of the three $g_1,g_2,g_3$ coupling constants in the context of strings
provided one of the following possible
alternatives is realized:
i) There is an intermediate grand unification
scale $M_X\sim 10^{16}$ GeV at which a GUT $simple$ group like
$SU(5)$ or $SO(10)$ is realized ; ii) There is instead some
$semi- simple$ group like $SU(4)\times SU(2)\times SU(2)$,
$SU(3)^3$ or $non$ $semi-simple$ group like $SU(5)\times U(1)$
beyond that scale. Both of these possibilities has its shortcomings.
The first requires that the Kac-Moody level of the gauge groups
$SU(5)$ or $SO(10)$ be bigger than one and the construction of
higher level models is both complicated and phenomenologically
problematic \KAC. Alternative ii) has the problem that there is further
relative running of the coupling  constants in the region
$M_X-M_{string}$  which will typically spoil the predictions
of the minimal susy model.
In any other possible alternative
it would be difficult to understand why the couplings tend to
join around  $10^{16}$ GeV, it would just be a mere
coincidence.
For example, it is possible to consider extensions of the minimal
particle content in such a way that the low-energy gauge
couplings directly meet around $10^{18}$ GeV.
This possibility was already considered in ref.\PLANCK\ .
In any case it is clear that
the present precision of the measurement of low energy
gauge couplings has reached a level which is sufficient to
test some fine details of string models.

\bigskip
\bigskip
We acknowledge useful  discussions with M. Cvetic, S. Ferrara, C. Kounnas
C. Mu\~noz, F. Quevedo, A.N. Schellekens and F. Zwirner.

\par \penalty-400 \vskip\chapterskip
   \spacecheck\referenceminspace \immediate\closeout\referencewrite
   \referenceopenfalse
   \line{\fourteenrm\hfil REFERENCES\hfil}\vskip\headskip
   \input referenc.texauxil
   
\endpage

\centerline{\bf Figure Captions}
\vskip1cm

\noindent{\bf Figure 1:}
$\sin^2\theta_W(M_Z)$ as a function of the compactification radius$^2$
${\rm Re}T=R^2$ for different values of $\delta A$. The shaded area
corresponds to the experimentally allowed range.

\noindent{\bf Figure 2:}
$\alpha_s(M_Z)$ as a function of
${\rm Re}T$ for different values of $\delta B$.

\vfill\eject\bye